\documentclass[useAMS,usenatbib]{mn2e}
\usepackage{graphicx,epsfig,amssymb,multirow,color}

\def\eg{{e.g.,~}}
\def\ie{{i.e.,~}}
\def\Ha{H$\alpha$ }
\def\Hb{H$\beta$ }
\def\Lya{Ly$\alpha$ }
\newcommand{\apj}{ApJ}%                                         % Journal abbreviations
\newcommand{\apjs}{ApJS}
\newcommand{\apjl}{ApJL}
\newcommand{\aap}{A{\&}A}

\newcommand{\mnras}{MNRAS}
\newcommand{\aj}{AJ}

\newcommand{\na}{New Astronomy}

\title[Consequences of bursty star formation]{Consequences of bursty star formation on galaxy observables at high redshifts}

\author[A. Dom\'inguez et al.] {Alberto Dom\'inguez$^{1,2}$\thanks{E-mail: alberto@clemson.edu}, Brian Siana$^{1}$, Alyson M.~Brooks$^{3}$, Charlotte R.~Christensen$^4$\newauthor Gustavo Bruzual$^{5}$, Daniel P.~Stark$^{6}$ and Anahita Alavi$^{1}$\\
	$^1$Department of Physics \& Astronomy, University of California Riverside, Riverside, CA 92521, USA\\
	$^2$Department of Physics \& Astronomy, Clemson University, Clemson, SC 29634, USA\\
	$^3$Department of Physics \& Astronomy, Rutgers, The State University of New Jersey, Piscataway, NJ 08854, USA\\
	$^4$Grinnell College Physics Department, Noyce Science Center, Grinnell, IA 50112, USA\\
	$^5$Centro de Radioastronom\'ia y Astrof\'isica, UNAM, Campus Morelia, Michoac\'an, 58089, Mexico\\
	$^6$Steward Observatory, University of Arizona, 933 N Cherry Ave, Tucson, AZ 85721, USA}

\begin{document}
\label{firstpage}

\date{\today}

\pagerange{\pageref{firstpage}--\pageref{lastpage}} \pubyear{2013}

\maketitle

\begin{abstract}
The star formation histories (SFHs) of dwarf galaxies are thought to be \emph{bursty}, with large -- order of magnitude -- changes in the star formation rate on timescales similar to O-star lifetimes.  As a result, the standard interpretations of many galaxy observables (which assume a slowly varying SFH) are often incorrect. Here, we use the SFHs from hydro-dynamical simulations to investigate the effects of bursty SFHs on sample selection and interpretation of observables and make predictions to confirm such SFHs in future surveys. First, because dwarf galaxies' star formation rates change rapidly, the mass-to-light ratio is also changing rapidly in both the ionizing continuum and, to a lesser extent, the non-ionizing UV continuum. Therefore, flux limited surveys are highly biased toward selecting galaxies in the \emph{burst} phase and very deep observations are required to detect all dwarf galaxies at a given stellar mass. Second, we show that a $\log_{10}[\nu L_{\nu}(1500{\rm \AA})/L_{{\rm H}\alpha}]>2.5$ implies a very recent quenching of star formation and can be used as evidence of stellar feedback regulating star formation. Third, we show that the ionizing continuum can be significantly higher than when assuming a constant SFH, which can affect the interpretation of nebular emission line equivalent widths and direct ionizing continuum detections. Finally, we show that a star formation rate estimate based on continuum measurements only (and not on nebular tracers such as the hydrogen Balmer lines) will not trace the rapid changes in star formation and will give the false impression of a star-forming main sequence with low dispersion.
\end{abstract}


\section{Introduction}
\label{sec:intro}
One of the most important results derived from deep galaxy surveys is that star formation in the more massive galaxies (galaxies with stellar masses larger than approximately $10^{9}$~M$_{\odot}$) is regulated by gradual processes such as gas exhaustion (\eg \citealt{noeske07b}). This fact implies that the star formation history (SFH) of these galaxies can be described by slowly varying functions of time (on timescales larger than approximately 100~Myr). However, stochastic processes are expected to dominate in dwarf galaxies (galaxies with stellar masses lower than approximately $10^{9}$~M$_{\odot}$). Specifically, the star formation in dwarf galaxies occurs in only a small number of regions and in a small volume. In such systems, feedback from supernovae can heat and expel gas from a volume comparable to the entire volume of the cold gas region, resulting in a temporary quenching of star formation. Therefore, the SFHs of dwarf galaxies are characterized by frequent bursts of star formation and subsequent quenching (on timescales of the order of a few Myr; \eg \citealt{shen13,hopkins13}). This burstiness may be caused mainly by supernovae feedback (\eg \citealt{governato12,teyssier13}). Bursty SFHs complicate our interpretation of observable properties, and oversimplification of these SFHs may lead to significant biases in determinations of fundamental galaxy properties (\eg \citealt{boquien14}).

Recent studies are showing that low mass galaxies at high redshift are contributing significantly to the total star formation rate density (\citealt{alavi14}). Understanding these low mass galaxies, especially at $z\sim 2$--3 when the star formation rate (SFR) of the Universe peaked (\citealt{reddy09}), is essential for explaining a number of phenomena in the early Universe.

First, it is thought that dwarf galaxies reionized the intergalactic medium at $z>7$ and provided a significant fraction of the ionizing background at $2<z<7$ (\citealt{haardt12,becker13}). Many investigations of escaping Lyman continuum from galaxies are necessarily conducted at redshifts of $2<z<3.5$ (\citealt{vanzella10,nestor13,mostardi13}), as the IGM becomes more opaque at higher redshift (\citealt{prochaska09}). In these studies, one typically assumes an intrinsic Lyman continuum flux based on the non-ionizing UV flux at approximately 1500~\AA~(\citealt{siana07}). However, this conversion usually assumes constant star formation for a long duration (longer than 100~Myr). The bursty SFHs of dwarf galaxies will significantly affect the level of Lyman continuum flux, the selection of the galaxies, and our interpretation of the global \emph{escape fraction} of Lyman continuum photons. Furthermore, the bursty star formation also has important implications for interpreting the byproducts of the Lyman continuum (\ie nebular emission lines) in dwarf galaxies.

Second, among more massive  galaxies, there is a tight correlation between star formation and stellar mass.  This correlation, called the star-forming main sequence, may exist from the local Universe up to $z\sim 6$ (\eg \citealt{brinchmann04,noeske07a,pannella09,whitaker12,speagle14}). These observations lead to the interpretation, mentioned above, that star formation in these galaxies is predominantly regulated by gradual processes. Whether this tight relation remains at lower masses is still unclear, although very recently progress has been made by \citet{whitaker14} in measuring the \emph{average} main sequence in lower mass galaxies. Bursty star formation should increase the dispersion in this relation. Unfortunately, the star formation indicators that we typically use, namely ultraviolet (UV) and infrared (IR) fluxes, vary on much larger time scales than the star formation in dwarf galaxies. It is therefore possible that we are not able to detect an increased scatter in the SFR at low mass with these traditional indicators.

In this paper, we analyze how bursty SFHs produced by hydro-dynamical simulations affect our current knowledge of the two issues stated above: the ionizing photon production and its subsequent effect on observables (\eg nebular line strengths, searches for escaping Lyman continuum), and the appearance of a tight star-forming main sequence. The methodology is based on using a stellar population synthesis code to model the spectral energy distributions (SEDs) of these bursty galaxies, which cover the stellar mass range from $10^{7}$~M$_{\odot}$ to $10^{10}$~M$_{\odot}$ at $z\sim 2$. Physical galaxy properties such as stellar masses and star-formation rates are also derived by using SED fitting, which are compared with their known values from the model galaxies (\eg \citealt{pacifici12}).

This paper is organized as follows. In \S\ref{sec:simul}, we briefly describe our hydro-dynamical simulations. Then, \S\ref{sec:modeling} gives details on the extraction of the galaxy SEDs from the results of the simulations. Later, \S\ref{sec:ion} shows the results from our analysis in terms of the ionizing photon production and discusses them. Then, \S\ref{sec:ms} analyzes the star-forming main sequence. In \S\ref{sec:sed}, we compare physical galaxy properties from the model SEDs and the SED fitting. Finally, we summarize in \S\ref{sec:summary} the main conclusions of our analysis. Throughout the paper, we use the same cosmology assumed by the simulations, this is, a $\Lambda$CDM cosmology with $H_{0}=73$~km~s$^{-1}$~Mpc$^{-1}$, $\Omega_{m}=0.24$, and $\Omega_{\Lambda}=0.76$ (according to data from the third release of the Wilkinson Microwave Anisotropy Probe, WMAP~3, \citealt{spergel07}).

\section{Galaxy simulations}
\label{sec:simul}
The bursty in-situ SFHs used in our analysis, defined as the star formation that occurs within the most massive progenitor of the main halo, are extracted from the hydro-dynamical simulations that are described in this section.

\subsection{Details of the simulations}
The high-resolution simulations used in this work were run using the \emph{volume re-normalization technique} (or \emph{zoom-in technique}) with the N-body plus smoothed-particle hydrodynamics code \texttt{GASOLINE} (\citealt{wadsley04}). The galaxies were selected to span a range of merger histories. All 11 of the galaxies in this work have been published and discussed more extensively in previous papers. Importantly for this work, earlier papers using these simulated galaxies have shown that they have realistic rotation curves (\citealt{christensen14}), that they match the $z=0$ stellar mass to halo mass relation derived from abundance matching techniques (\citealt{munshi13}), and that the bursty star formation history in the seven lowest mass galaxies leads to the creation of dark matter cores in their central density profiles (\citealt{governato12}). These simulations also match the observed mass--metallicity relation (\citealt{brooks07}), the sizes of galaxy disks (\citealt{brooks11}), and result in a realistic population of satellites in Milky Way-mass galaxies (\citealt{zolotov12,brooks14}). These simulated galaxies are among the highest resolution achieved in cosmological simulations to date. The high resolution allows us to resolve the high density gas clouds where stars form.  With this scheme, we have successfully matched a large range of observed galaxy properties, as mentioned above. An extensive discussion of the star formation and feedback scheme in these simulations is found in \citet{christensen12}. Below, we summarize the salient information about the simulations. 

The spline force softening in these simulations ranges from 65~pc (h2003) to 87~pc (h516 and h799) to 174~pc (all others). High resolution dark matter particles have masses of 6661~M$_{\odot}$ (1.3$\times$10$^{5}$~M$_{\odot}$), while gas particles start with 1407~M$_{\odot}$ (2.7$\times$10$^{4}$~M$_{\odot}$) for the lowest (largest) mass galaxies.  Star particles are born with 30\% of the mass of their parent gas particle and lose mass through supernova (both Ia and II) and stellar winds. Each of these galaxies has between one to five million dark matter particles within the virial radius at $z=0$. The simulations not only include metal line cooling and metal diffusion (\citealt{shen10}), but a prescription for the formation (both gas-phase and on dust grains) and destruction (primarily photo-dissociation by Lyman-Werner radiation from nearby stellar populations) of H$_2$ (\citealt{christensen12}). Star formation is tied directly to the presence of H$_2$, as observed (\citealt{leroy08,bigiel08,blanc09,bigiel10,schruba11}). These simulations also include a uniform UV background that turns on at $z=9$, mimicking cosmic reionization following a modified version of \citet{haardt01}.  

\begin{table}
\centering
%\begin{tabular}{|l|c|c|c|}
\begin{tabular}{|l|c|c|c|}
\hline
%Simulation & $\log_{10}({\rm M}_{*}^{BC03}/{\rm M}_{\odot})$ & \multicolumn{2}{|c|}{$\log_{10}({\rm M}_{*}^{sim}/{\rm M}_{\odot})$}\\
Simulation & \multicolumn{2}{|c|}{$\log_{10}({\rm M}_{*}/{\rm M}_{\odot})$}\\
\hline
%& $z=2$ & $z=2$ & $z=0$\\
& $z=2$ & $z=0$\\
\hline
\hline
h2003.grp1 & 6.97 & 7.05\\
h799.grp1 & 7.07 & 7.64\\
h516.grp1 & 7.70 & 8.05\\
h986.grp3 & 8.29 & --\\
h603.grp1 & 8.35 & 9.45\\
h986.grp2 & 8.41 & --\\
h986.grp1 & 8.51 & 9.29\\
h239.grp1 & 9.09 & 10.25\\
h285.grp1 & 9.33 & 10.29\\
h258.grp1 & 9.43 & 10.22\\
h277.grp1 & 9.94 & 10.33\\
\hline
\end{tabular}
\caption{The first column lists the simulation name. The second and third columns show the stellar masses at $z=2$ and $z=0$ (respectively; see text for details) from the in-situ SFHs output by the \texttt{GALAXEV} code. The galaxies are sorted increasingly from top to bottom by their stellar mass at $z=2$. The h986.grp2 and h986.grp3 merge with the most massive progenitor h986.grp1 by $z=0$.}
\label{tab1}
\end{table}

Critically for this work, the force resolution of these runs (65--174~pc) allows high density regions where stars form to be resolved ($\rho \gtrsim100$ amu/cm$^3$, comparable to the average density in giant molecular clouds). Star formation in the simulations is done in 1~Myr increments.  That is, every 1~Myr we identify the gas that is cold and dense enough to form stars. For all gas particles that meet this criterion, there is a $10\%\times f_{{\rm H}2}$ chance (where $f_{{\rm H}2}$ is the molecular fraction of that gas particle) that they will spawn a star particle. The high resolutions allow for a physically motivated stellar and supernova feedback scheme that injects energy locally back into the interstellar medium (ISM, following \citealt{stinson06}) rather than the adoption of an analytic prescription to model feedback globally. This local injection of energy can heat the surrounding gas such that star formation is completely shut down for a brief period (\citealt{stinson07}). In low mass dwarfs, this leads to a bursty SFH.

We follow each galaxy back to high $z$, identifying at each output timestep the halo progenitor that contains most of the stellar mass at lower $z$. We identify halos at all steps using the Amiga Halo Finder (\citealt{gill04,knollmann09}). Two of the halos (h986.grp2 and h986.grp3) merge with h986.grp1 at $z<1$, and no longer exist as independent galaxies at $z=0$.

There are other authors who simulate galaxies using different ISM/feedback recipes but still obtain bursty SFHs similar to ours (\citealt{shen13,hopkins13,ceverino14,trujillo-gomez14}). Relevant to this paper is the fact that, as galaxy stellar mass decreases, the dispersion in star formation rate increases, especially on short time scales ($<100$~Myr). We have calculated the dispersion of SFRs for our simulated galaxies on different time scales and find that the trend is similar to the values shown in Figure~12 of \citet{hopkins13}. Therefore, the conclusions in our paper should broadly be true in other simulations as well.

\begin{figure}
\includegraphics[width=\columnwidth]{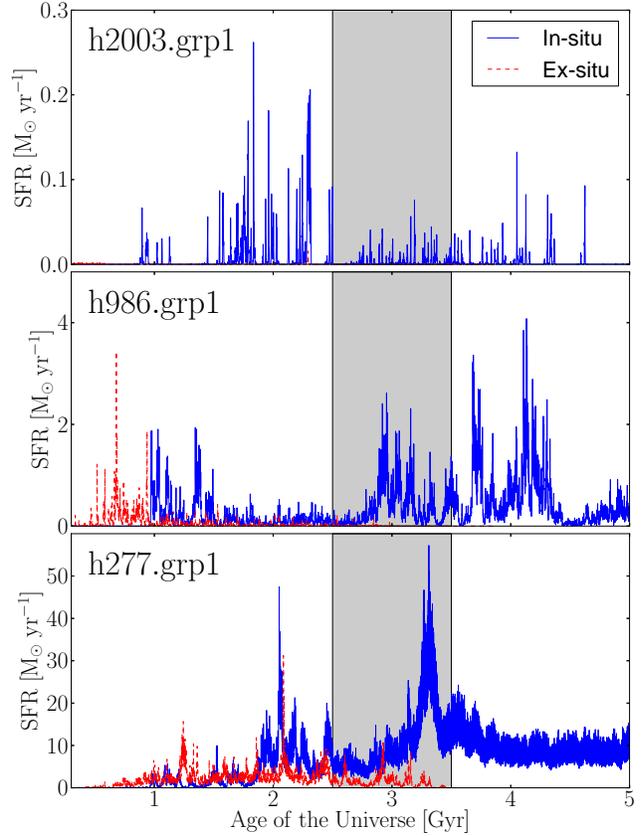}
\caption{The star formation history of three of our simulated galaxies. Galaxies are shown, from top to bottom, in order of increasing stellar mass at $z=2$. These stellar masses are $\log_{10}({\rm M}_{*}/{\rm M}_{\odot})=6.97$, 8.51, and 9.94. The blue line is the in-situ star formation, whereas the red line shows the ex-situ star formation that becomes part of the galaxy by $z\sim 2$. The shaded region shows the time range where the galaxy SEDs are extracted, when the Universe is between 2.5 and 3.5~Gyr old. Given our cosmology, these ages correspond to $z=2.74$ and $z=1.97$, respectively. The lower mass galaxies have more dramatic changes in SF on short timescales, whereas the more massive galaxies have smaller fractional variations, and on longer timescales}
\label{fig:SFHs}
\end{figure}

\subsection{Simulated star formation rate histories}
Figure~\ref{fig:SFHs} shows the in-situ SFH of three of our 11 simulated galaxies. The SFHs of the lower mass galaxies are characterized by complicated functions of time featuring short bursts of star formation. These bursts can reach an order of magnitude larger star formation rate than a rather quiescent state in timescales of a few Myr or even less. Figure~\ref{fig:SFHs} also shows the ex-situ SFHs of the stars that are in the galaxies at $z\sim 2$. These are stars formed outside the main halo that through mergers become part of the main galaxy by $z\sim 2$. We stress that, as seen in Figure~\ref{fig:SFHs}, the ex-situ contribution is rather low compared to the in-situ star formation, particularly in the lower mass galaxies that are the main targets of this study. Observational results from other authors such as \citet{behroozi13} also suggest that the vast majority of all stars in low mass halos (by $z\sim 2$) are formed in situ.

\begin{figure}
\includegraphics[width=\columnwidth]{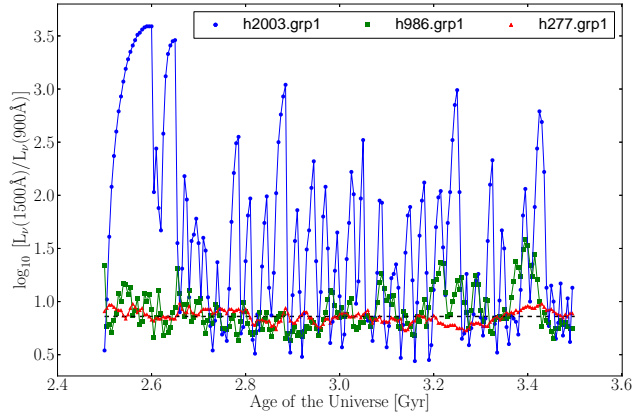}
\caption{The ratio between the continuum luminosities $\log_{10}[L_{\nu}(1500{\rm \AA})/L_{\nu}(900{\rm \AA})]$ as a function of age for three galaxies of significantly different stellar mass. These galaxies are our lowest mass galaxy h2003.grp1 ($\log_{10}({\rm M}_{*}/{\rm M}_{\odot})=6.97$), the intermediate mass galaxy h986.grp1 ($\log_{10}({\rm M}_{*}/{\rm M}_{\odot})=8.51$), and our largest mass galaxy h277.grp1 ($\log_{10}({\rm M}_{*}/{\rm M}_{\odot})=9.94$). The horizontal line marks the commonly assumed ratio of $\log_{10}[L_{\nu}(1500{\rm \AA})/L_{\nu}(900{\rm \AA})]=0.84$ (\citealt{siana07}). The repeated quenching of star formation in low mass galaxies causes the 900~\AA\ continuum luminosity to drop dramatically, resulting in large variations in the ratio.}
\label{fig:UVratio}
\end{figure}

\section{Modeling of the galaxy spectral energy distributions}
\label{sec:modeling}
We use the stellar population synthesis code \texttt{GALAXEV} by \citet[][hereafter BC03]{bruzual03} to determine the galaxy SEDs and stellar masses from the simulated SFHs. We note that BC03 accurately traces the mass loss from supernovae. The BC03 stellar masses of our galaxies are reported in Table~\ref{tab1} for $z=2$ and $z=0$. These galaxies span the mass range from roughly 10$^{7}$--10$^{10}$~M$_{\odot}$ at $z\sim 2$. Our model SEDs are extracted at times when the Universe is from 2.5~Gyr to 3.5~Gyr old in steps of 5~Myr (a total of 201 SEDs for each simulated SFH). These ages correspond to redshifts from $z=2.74$ to $z=1.97$ according to the cosmology of the simulations. Though we only output the SEDs at 5~Myr intervals, it is important to note that we are using SFHs with 1~Myr resolution. This is necessary for capturing star formation changes on short time scales. The initial mass function (IMF) is assumed to be given by \citet{chabrier03} and the metallicity is fixed to $Z=0.2 Z_{\odot}$, where $Z_{\odot}$ is solar metallicity. This metallicity is roughly consistent with the sub-solar metallicities measured in $\log_{10}({\rm M}_{*}/{\rm M}_{\odot}) < 8$ at $z\sim 2$ (\citealt{belli13,henry13,sanders14}). No dust extinction is included. We neglect secondary effects of metallicity and dust to isolate the effects of bursty star formation on observable quantities.

From the model SEDs, we derive the UV luminosity in the continuum at two different wavelengths: the ionizing continuum at 900~\AA~(also known as Lyman continuum, or LyC) and the non-ionizing continuum at 1500~\AA, where most observational studies easily detect high redshift galaxies.

The BC03 code outputs the number of ionizing photons per second ($\dot{N}$), which is used to estimate the \Hb emission-line luminosity ($L_{{\rm H}\beta}$, in erg/s) as,

\begin{equation}
L_{{\rm H}\beta}=4.757\times 10^{-13} \dot{N}%\hspace{3cm}{\rm [erg/s]}
\end{equation}

\begin{figure}
\includegraphics[width=\columnwidth]{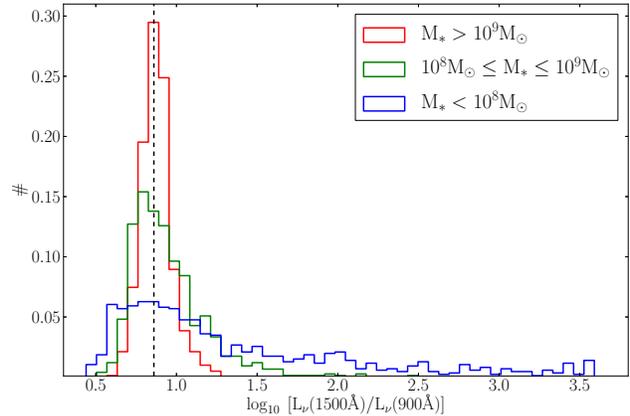}
\caption{The normalized $L_{\nu}(1500{\rm \AA})/L_{\nu}(900{\rm \AA})$ distributions in three stellar mass bins. The vertical line marks the commonly assumed ratio of $\log_{10}[L_{\nu}(1500{\rm \AA})/L_{\nu}(900{\rm \AA})]=0.84$ (\citealt{siana07}). Low mass galaxies have a larger distribution of $L_{\nu}(1500{\rm \AA})/L_{\nu}(900{\rm \AA})$, with larger ratios shortly after star formation quenching and lower ratios immediately after a new burst of star formation.}
\label{fig:histUV}
\end{figure}

\noindent (see \citealt{krueger95}). This luminosity is necessary as a reference from which to derive other emission-line luminosities. In our case, we consider \Lya and H$\alpha$, whose ratios are $L_{{\rm Ly}\alpha}/L_{{\rm H}\beta}=22.20$ and $L_{{\rm H}\alpha}/L_{{\rm H}\beta}=2.87$ (case B recombination; \citealt{osterbrock06}). We assume that all of the ionizing photons are absorbed by hydrogen in the ISM of galaxies - that is, that the LyC escape fraction is near zero.  Although this choice may not be realistic for a subset of galaxies (\eg \citealt{nestor13,mostardi13}), only very large escape fractions (\ie larger than approximately 0.5) would significantly affect our results. The rest-frame \Lya and \Ha equivalent widths (EWs) are calculated from the model SEDs as the ratio between the luminosity of the line and the continuum luminosity per unit of wavelength at the central wavelength of the line. We do not apply any correction to account for stellar absorption in H$\alpha$. However, this choice will not substantially affect our results since \Ha in absorption is typically small in young galaxies (\eg \citealt{dominguez13}).

As we mentioned in \S\ref{sec:simul}, a small fraction of the stellar mass is formed ex-situ and becomes part of the galaxy through mergers. Importantly for the subsequent analysis, the stellar mass produced ex-situ will not contribute to the LyC or UV photon emission since the star formation occurred long enough before coalescence.

\section{Study of the ionizing photon production}
\label{sec:ion}
The UV and \Ha luminosities are directly related to star formation. The former, $L_{\nu}(1500{\rm \AA})$ or $L_{UV}$, is dominated by light coming from young and massive O and B stars ($\gtrsim 3$~M$_{\odot}$). This indicator is sensitive to the recent SFR given the lifetime of the massive stars that produce it ($\lesssim 100$~Myr). The latter, $L_{{\rm H}\alpha}$, is produced by ionizing radiation coming from nebular regions heated by extremely massive stars ($\gtrsim 20$~M$_{\odot}$). This indicator is sensitive to the instantaneous SFR since it is dominated by stars with very short lifetime ($\lesssim 5$~Myr). According to BC03 models, after an instantaneous burst of star formation, it takes only 5~Myr for $L_{\nu}(900{\rm \AA})$ to decrease an order of magnitude, whereas it takes 30~Myr for $L_{\nu}(1500{\rm \AA})$. Results of our analysis in terms of the production of ionizing radiation and nebular emission lines are discussed in this section.

\begin{table}
\centering
\begin{tabular}{|l|c|c|c|}
\cline{2-4}
& \multicolumn{3}{c}{Stellar Mass Range [M$_{\odot}$]}\\
\hline
Standard deviation & $\leq 10^{8}$ & $10^{8}-10^{9}$ & $\geq 10^{9}$\\
\hline
$\log_{10}~[{\rm L}_{\nu}(1500{\rm \AA})/{\rm L}_{\nu}(900{\rm \AA})]$ & 0.74 & 0.23 & 0.09\\
$\log_{10}~({\rm Ly}\alpha~{\rm EW}/{\rm \AA})$ & 0.53 & 0.18 & 0.07\\
$\log_{10}~({\rm H}\alpha~{\rm EW}/{\rm \AA})$ & 0.97 & 0.39 & 0.20\\
$\log_{10}~({\rm SFR}/{\rm M}_{\odot}~{\rm yr}^{-1})$ & $>$0.65 & $>$0.47 & 0.26\\
$\log_{10}~[L_{\nu}(900{\rm \AA})/{\rm erg~s}^{-1}~{\rm Hz}^{-1}]$ & 1.03 & 0.45 & 0.22\\
$\log_{10}~(L_{{\rm H}\alpha}/{\rm erg~s}^{-1})$ & 1.03 & 0.45 & 0.22\\
$\log_{10}~[L_{\nu}(1500{\rm \AA})/{\rm erg~s}^{-1}~{\rm Hz}^{-1}]$ & 0.43 & 0.28 & 0.16\\
$\log_{10}~[L_{\nu}(5500{\rm \AA})/{\rm erg~s}^{-1}~{\rm Hz}^{-1}]$ & 0.15 & 0.11 & 0.06\\
$\log_{10}~[L_{\nu}(1.2\mu{\rm m})/{\rm erg~s}^{-1}~{\rm Hz}^{-1}]$ & 0.10 & 0.08 & 0.05\\\hline
\end{tabular}
\caption{The standard deviation of the quantities listed in column 1 in three different stellar mass bins after removing the stellar mass trend. We are reporting only lower limits for the two lower bins of the SFR because the values for $SFR=0$ are not included in the standard deviation calculation.}
\label{tab2}
\end{table}

\subsection{Ionizing photon production}
The SFRs of dwarf galaxies are expected to change on time-scales similar to lifetimes of the ionizing O-stars. Therefore, we might expect wide variations in the ionizing photon production rate in these systems.

The ratio between the continuum luminosities $L_{\nu}(1500{\rm \AA})/L_{\nu}(900{\rm \AA})$ is typically used to derive the relative escape fraction of ionizing photons and also to convert from observed UV luminosity functions to LyC photon production rates (\eg \citealt{steidel01,siana07,siana10}). This flux ratio is usually assumed as a constant ratio of approximately 7 (or 0.84 in $\log_{10}$). This value is reached when the equilibrium state is produced under continuous star formation when the number of stars producing LyC and 1500~\AA~flux is constant (or about 200~Myr after the burst). The $\log_{10}[L_{\nu}(1500{\rm \AA})/L_{\nu}(900{\rm \AA})]$ is shown in Figure~\ref{fig:UVratio} as a function of time for three different galaxies. We see that the scatter of $\log_{10}[L_{\nu}(1500{\rm \AA})/L_{\nu}(900{\rm \AA})]$ is significantly larger in the low mass galaxies, varying by nearly an order of magnitude on short time scales but only around 0.1~dex for the larger mass galaxies.

In Figure~\ref{fig:histUV}, we show the normalized distribution of $\log_{10}[L_{\nu}(1500{\rm \AA})/L_{\nu}(900{\rm \AA})]$ for three different stellar mass bins that include all evolutionary stages. From the analysis of our simulations, it is clear that the assumption of $\log_{10}[L_{\nu}(1500{\rm \AA})/L_{\nu}(900{\rm \AA})]\sim 0.84$ is valid for the larger mass galaxies but not for the lower mass galaxies due to significantly large scatter. The scatter is quantified as the standard deviation of the distribution in three stellar mass bins in Table~\ref{tab2}.

\begin{figure}
\includegraphics[width=\columnwidth]{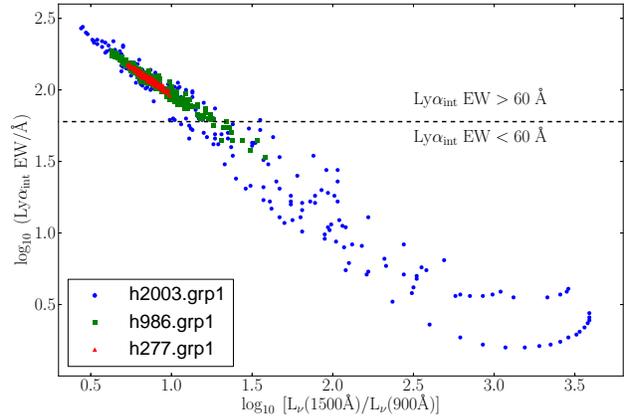}
\caption{The intrinsic \Lya EW versus $L_{\nu}(1500{\rm \AA})/L_{\nu}(900{\rm \AA})$ for three galaxies of significantly different stellar mass at different evolutionary stages. These galaxies are our lowest mass galaxy h2003.grp1 ($\log_{10}({\rm M}_{*}/{\rm M}_{\odot})=6.97$), the intermediate mass galaxy h986.grp1 ($\log_{10}({\rm M}_{*}/{\rm M}_{\odot})=8.51$), and our largest mass galaxy h277.grp1 ($\log_{10}({\rm M}_{*}/{\rm M}_{\odot})=9.94$). The evolutionary stages above the horizontal line are considered typical \Lya emitters. In general, non-\Lya emitter are significantly faint in the LyC. Though many LyC searches of high redshift galaxies are not targeting low mass galaxies with low \Lya EW, those galaxies likely do not have strong Lyman continuum luminosities anyway, as the star formation may have recently been quenched. For instance, our galaxy of lowest mass is a \Lya emitter for 57\% of the time.}
\label{fig:LyavsUV}
\end{figure}

This scatter in $L_{\nu}(1500{\rm \AA})/L_{\nu}(900{\rm \AA})$ in low mass galaxies will require particular care in interpreting studies of escaping LyC. First, there are some galaxies, caught just when a burst begins, that have large LyC luminosities relative to the non-ionizing 1500~\AA\ luminosities. This will make them easier to detect in direct LyC searches. These same galaxies will have very high intrinsic (unextinguished by dust) \Lya EW (see Figure \ref{fig:LyavsUV}). Therefore, galaxies selected for high \Lya EW (\eg \citealt{nestor11,mostardi13}) will have a higher likelihood of LyC detection than galaxies with lower \Lya EW, even if the LyC escape fractions are similar. Second, low mass galaxies that have recently turned off their star formation will still be reasonably luminous at 1500~\AA, but will not be producing significant LyC photons. In Figure~\ref{fig:histUV}, we can see that 38\% of low mass (M$_{*} < 10^8$~M$_{\odot}$) galaxies have $\log_{10}[L_{\nu}(1500{\rm \AA})/L_{\nu}(900{\rm \AA})]>1.3$ (or above 20 in linear terms). These galaxies would often result in non-detections in direct LyC surveys, even if the LyC escape fractions of these galaxies were high. Ultimately, the \emph{burstiness} of star formation as a function of mass will need to be better determined and incorporated into any analysis of LyC observations. We note that in larger mass galaxies (M$_{*} > 10^9 M_{\odot}$), it is reasonable to assume a constant $L_{\nu}(1500{\rm \AA})/L_{\nu}(900{\rm \AA})$ because the star formation does not typically change much on short time scales. 

\citet{nestor11,nestor13} find a high ionizing emissivity from \Lya emitters at $z=3.09$, possibly explaining the entire ionizing background at that redshift. This finding is initially surprising because only a subset of UV-bright galaxies (Ly$\alpha$ emitters) is enough to explain, at least, a significant fraction of the ionizing background. However, it may be possible that Ly$\alpha$-emitting galaxies are the only ones producing significantly strong LyC. Indeed, this is shown in Figure~\ref{fig:LyavsUV}, where we plot \Lya EW as a function of $\log_{10}[L_{\nu}(1500{\rm \AA})/L_{\nu}(900{\rm \AA})]$. The horizontal line shown in Figure~\ref{fig:LyavsUV} defines typical \Lya emitters as intrinsic \Lya EW larger than 60~\AA~if we assume a \Lya escape fraction of 50\%. (Typically \Lya emitters are selected at \Lya EW larger than 30~\AA~and have \Lya escape fractions of 30--60\%, \citealt{gronwall07,ouchi08,nilsson09}.) As we see in the figure, in general, galaxies that are not \Lya emitters are by definition faint in the LyC. We stress here that all of these \Lya EW predictions are {\it intrinsic} measures before scattering by H{\sc i} and absorption by dust. Of course, most of the more massive, dusty galaxies, will not be \Lya emitters. Therefore, galaxies selected as \Lya emitters may be primarily composed of lower mass galaxies (with lower columns of neutral hydrogen and dust) in a burst phase. Determining exactly when each galaxy will be a Ly$\alpha$-emitter requires more detailed dust modeling and ray tracing, which is beyond the scope of this paper. Nonetheless, we stress that low mass galaxies may not produce significant LyC about half of the time and selection on high \Lya EW will miss these galaxies. Thus, it may not be surprising that a large fraction of the known ionizing background at high redshift is caused by low mass galaxies selected as Ly$\alpha$-emitters.

\begin{figure}
\includegraphics[width=\columnwidth]{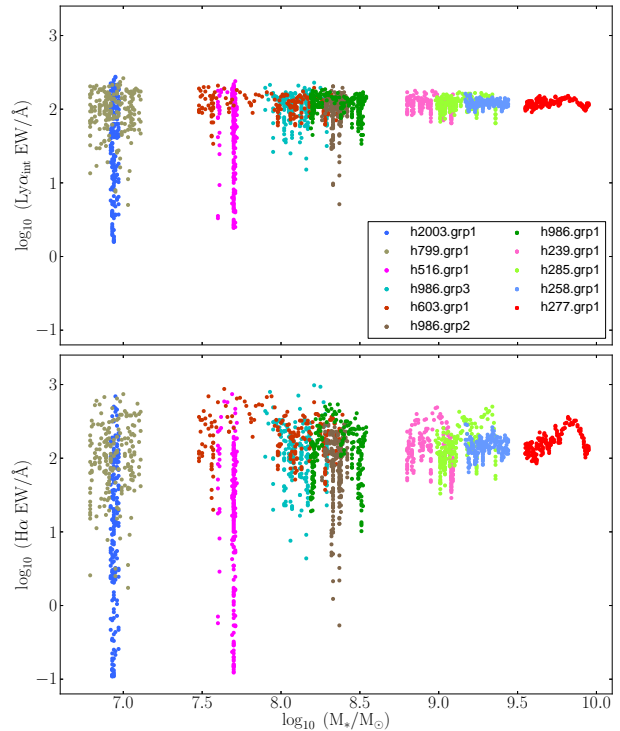}
\caption{The intrinsic \Lya (upper panel) and \Ha EW (lower panel) versus stellar mass for our sample of 11 galaxies. Each galaxy is represented by a different color. A circle is plotted every 5 Myr from times of 2.5 to 3.5~Gyr, which makes it a total of 201 data points for each galaxy. We note that the range of the $y$-axis is the same for both panels. The simulations predict a large population of low mass galaxies with \Ha EW lower than 30~\AA.}
\label{fig:EWs}
\end{figure}

\begin{figure}
\includegraphics[width=\columnwidth]{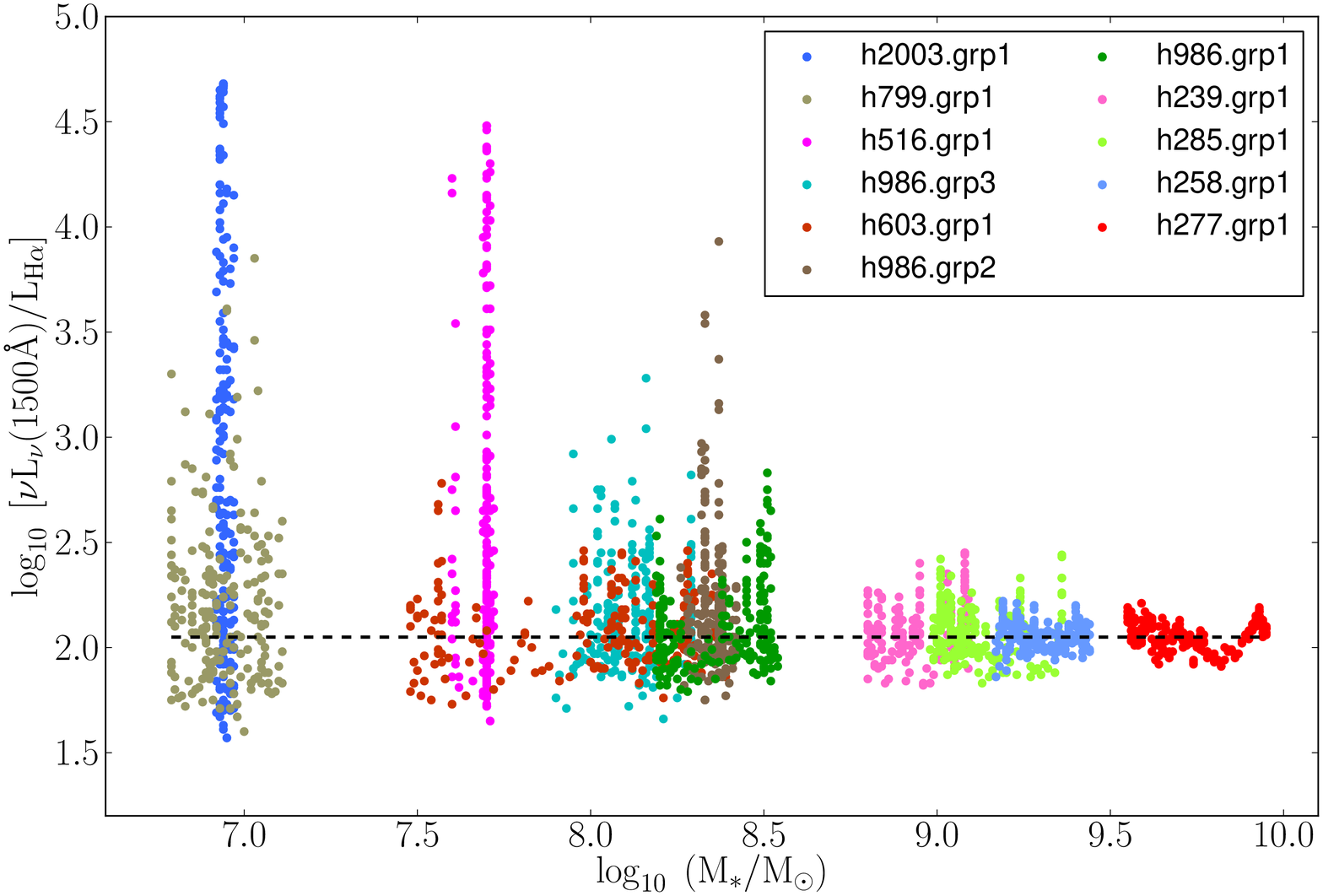}
\caption{The $\log_{10}[\nu L_{\nu}(1500{\rm \AA})/L_{{\rm H}\alpha}]$ versus stellar mass for our galaxies. The dashed line shows that ratio at 2.05 for a constant SFH after a steady state is reached. The bursty SFHs of the numerical simulations predict that a population exists with $\log_{10}[\nu L_{\nu}(1500{\rm \AA})/L_{{\rm H}\alpha}]>2.5$ at $\log_{10}({\rm M}_{*}/{\rm M}_{\odot})<8.5$, a clear indicator of rapid quenching of star formation.}
\label{fig:LUVHa}
\end{figure}

A potential problem in our calculations is the fact that \Lya photons significantly scatter with neutral hydrogen within the galaxy before escaping. This effect may produce a delay in the escape of the photons from the galaxy, which could smooth the \Lya luminosity time variation. However, given the size of a typical dwarf galaxy at $z\sim 2$ (of the order of a kpc) any delay due to a large number of scatters should be significantly smaller than the 1~Myr resolution of our simulations.

\subsection{Nebular emission-line equivalent width}
\label{sec:ews}
Recently, very young galaxies have been found at $z>1$ by identifying galaxies with large EW nebular emission lines (\citealt{vanderwel11,atek11,atek14,stark14}). In particular, the \Ha EW is a proxy of the specific star-formation rate of a galaxy (\ie the star-formation per unit of stellar mass). The rest-frame intrinsic \Lya and \Ha EWs are plotted in Figure~\ref{fig:EWs} as a function of stellar mass. Both EWs have a similar behaviour. Their scatter increases in the low mass galaxies as a consequence of the burstiness of the SFHs.

Table~\ref{tab2} quantifies the dispersion of the \Lya and \Ha EWs distribution as the standard deviation in three stellar mass bins. We note that the dispersion of the \Ha EW is larger than the dispersion of \Lya EW at all masses. The reason is that when the LyC is high (during a burst of star formation), both \Lya and \Ha emission-line luminosities increase by the same factor. However, the rest-UV continuum at 1216~\AA\ also increase strongly at the same time, whereas the continuum near 6563~\AA\ does not change as much as the rest-UV continuum. Therefore, the \Ha EW change is more dramatic.

The equivalent widths increase immediately following a new burst of star formation, as the ionizing photon production rate reacts immediately, but the continuum flux takes a longer time to increase. Conversely, a few Myr after the star formation turns off, the LyC disappears but the 1500~\AA\ flux fades much more slowly. Thus, the EWs quickly go to zero in recently quenched systems. We note that the plotted \Lya EWs are maximum values because it is expected that a significant fraction of \Lya photons will be absorbed by dust as it scatters through the interstellar and circumgalactic media.

\begin{figure*}
\includegraphics[width=18cm]{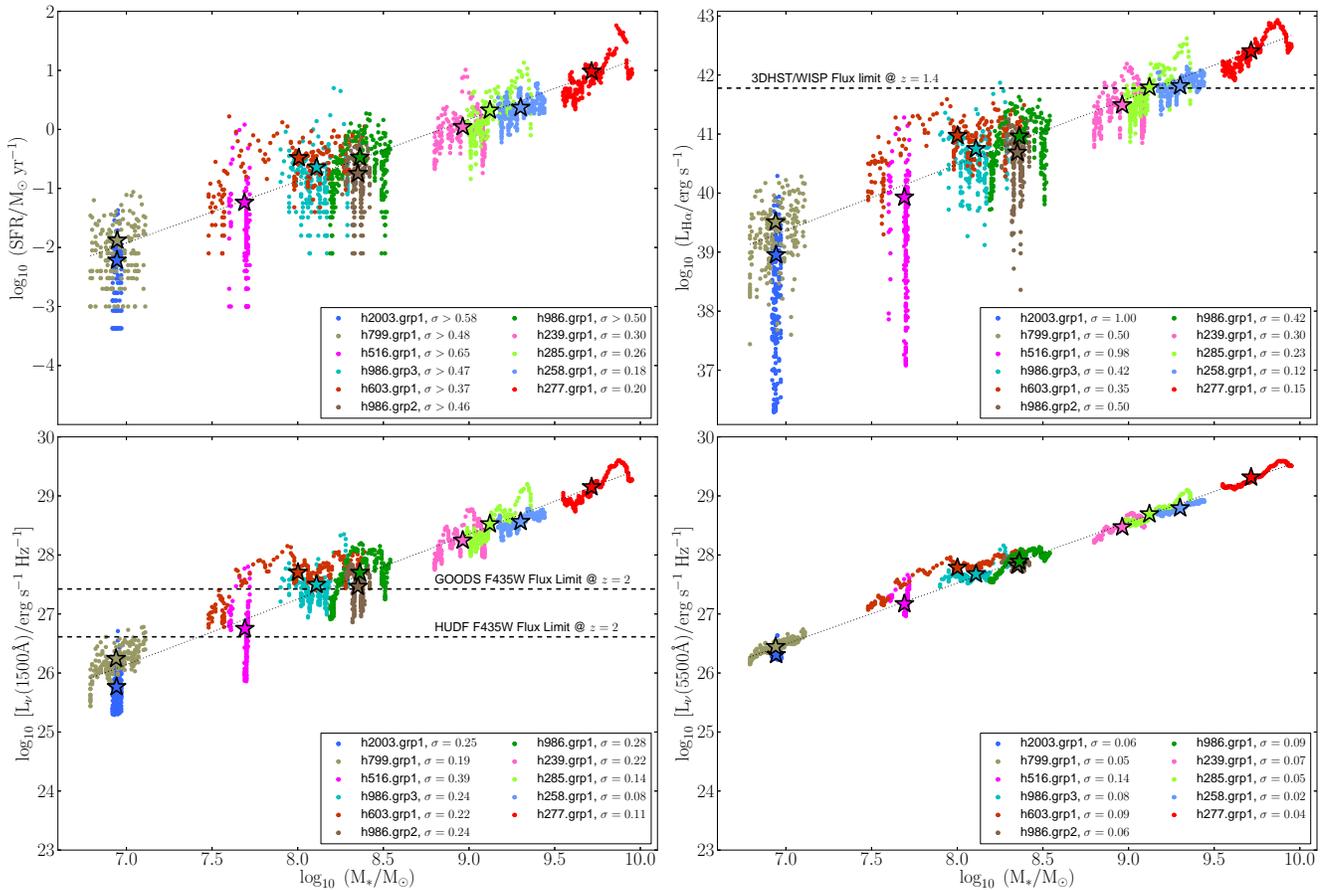}
\caption{\emph{(Top left panel)} The star formation rate from our simulations versus stellar mass. The best-fit straight line for the model galaxies (dotted line) is given by $\log_{10}({\rm SFR}/{\rm M}_{\odot}~{\rm yr}^{-1})=(1.05\pm 0.35)\times \log_{10}({\rm M}_{*}/{\rm M}_{\odot})-(9.25\pm 2.91)$. The scatter of the distributions (after removing the stellar mass trend) are given in the legends. Note that the evolutionary stages when the star formation rate of the galaxy, given by the simulations, is zero cannot be plotted. The stars are average values calculated as explained in the text. Note that all the panels show a range of 7 orders of magnitude in the $y$-axis. \emph{(Top right panel)} The \Ha luminosities versus stellar mass. The best fit straight line for the model galaxies (dotted line) is given by $\log_{10}(L_{{\rm H}\alpha}/{\rm erg}~{\rm s}^{-1})=(1.12\pm 0.35)\times \log_{10}({\rm M}_{*}/{\rm M}_{\odot})+(31.52\pm 2.90)$. The flux limit at $z=1.4$ from the WISP/3DHST galaxy survey (\citealt{atek10,brammer12}) is shown with a dashed line. \emph{(Bottom left panel)} The UV luminosities versus stellar mass. The best-fit straight line for the model galaxies (dotted line) is given by $\log_{10}(L_{UV}/{\rm erg}~{\rm s}^{-1}~{\rm Hz}^{-1})=(1.11\pm 0.35)\times \log_{10}({\rm M}_{*}/{\rm M}_{\odot})+(18.40\pm 2.90)$. The flux limits at $z=2$ from the GOODS ($B=27.80$, 5$\sigma$, 0.2'' radius, \citealt{giavalisco04}) and HUDF ($B=29.70$, 5$\sigma$, 0.2'' radius, \citealt{beckwith06}) galaxy surveys are shown with dashed lines. \emph{(Bottom right panel)} The optical luminosities versus stellar mass. The best-fit straight line for the model galaxies (dotted line) is given by $\log_{10}[L(5500~{\rm \AA})/{\rm erg}~{\rm s}^{-1}~{\rm Hz}^{-1}]=(1.03\pm 0.35)\times \log_{10}({\rm M}_{*}/{\rm M}_{\odot})+(19.25\pm 2.90)$. The slope of these relations are compatible with 1. Because the \Ha responds on short timescales, it accurately traces the rapidly varying SFR and has a similar scatter to the star-forming main sequence. The dispersion in $L_{UV}$ also increases at lower mass, but not with the same amplitude of the SFR. Finally, the $L_{\nu}(5500{\rm \AA})$ luminosity does not vary at all on the short time scales of the SFR variations.}
\label{fig:ms}
\end{figure*}

Two of the most common star formation rate indicators at high redshift are the UV and \Ha luminosities. However, as seen above, the ionizing continuum responsible for producing \Ha can change on very short time scales but the UV continuum takes longer to react. Therefore, we expect that the UV and H$\alpha$-derived SFRs may differ significantly depending on when the galaxy is observed (\eg \citealt{glazebrook99}). Figure~\ref{fig:LUVHa} shows the $\log_{10}[\nu L_{\nu}(1500{\rm \AA})/L_{{\rm H}\alpha}]$ as a function of stellar mass. The $\log_{10}[\nu L_{\nu}(1500{\rm \AA})/L_{{\rm H}\alpha}]$ tends to a value of 2.05 for the larger mass galaxies. This value corresponds to luminosities from a galaxy with constant SFR when the equilibrium in the number of O-stars to UV-emitting stars is reached. We can see in Figure~\ref{fig:LUVHa} that dwarf galaxies with bursty SFHs are characterized by $\log_{10}[\nu L_{\nu}(1500{\rm \AA})/L_{{\rm H}\alpha}]>2.5$ shortly after star formation has been quenched. Both $L_{\nu}(1500{\rm \AA})$ and $L_{{\rm H}\alpha}$ are typically observed by deep galaxy surveys. Therefore, this figure provides a reliable test for bursty SFHs in star-forming galaxies based on large ratios of the UV continuum at 1500~\AA\ and the \Ha emission luminosity. The effect of periodic bursty SFHs on the ratio of UV and \Ha luminosity is studied in low redshift dwarf galaxies by some authors such as \citet{iglesias-paramo04}, \citet{boselli09}, \citet{meurer09} and \citet{weisz12}. These authors use analytical models to explore the amplitude, duty cycle, and duration of bursts of star formation by fitting the observed distribution of UV to \Ha ratios. In our analysis, we take an alternative approach using SFHs motivated by results from hydro-dynamical cosmological simulations.

\subsection{Populating the IMF stochastically}
At low SFRs, additional analysis uncertainties may arise due to poor sampling of the high mass end of the IMF. Individual bursts may also have different IMFs. Stochastic effects on the IMF have been studied by other authors such as \citet{fumagalli11}, \citet{forero-romero13}, and \citet{dasilva14} using the code Stochastically Lighting Up Galaxies by \citet{dasilva12}. According to \citet[][see their Figure 1 and Figure 2]{forero-romero13} this effect becomes important at star formation rates lower than 0.01~M$_{\odot}$yr$^{-1}$. From the examples shown in our Figure~\ref{fig:SFHs}, we note that the stellar mass involved in the bursts considered is sufficiently high that individual bursts fairly sample the underlying IMF. This implies that this stochasticity effect should not be too large in our sample. Though we expect this to be a minor effect, it will result in a slight decrease in the distribution of \Lya and \Ha EWs at the low mass end in Figure~\ref{fig:EWs}. However, all the main observational conclusions of this paper will not be affected.

\section{The star-forming main sequence}
\label{sec:ms}
Typically, when evaluating trends in SFHs, the SFR is plotted as a function of stellar mass to show the star-forming main sequence. Our theoretical work complements other observational analyses of the main sequence. Here, we choose to analyze directly the luminosities instead of the SFRs to avoid any uncertainties from their conversion.

In Figure~\ref{fig:ms}, we plot the SFR versus stellar mass. In addition, three other observables are plotted in Figure~\ref{fig:ms}, H$\alpha$ luminosity, UV luminosity density, and optical luminosity density versus stellar mass. One can see that the large variation in SFR in low mass galaxies is traced very well by \Ha as it reacts quickly to the changing SFR. The UV variations are not as large as the SFR variations, and the optical luminosity is essentially unaffected by the SFR variations.  

Table~\ref{tab2} lists the standard deviation of these observables in three mass bins. At large mass, the dispersion in SFR is small, and both \Ha and UV reasonably trace the SFR variations (though \Ha is certainly better). However, at lower masses (M$_{*}<10^{9}$~M$_{\odot}$), as the SFR variations increase, the UV dramatically under-predicts the SFR variation. Therefore, our results show that, for low mass galaxies, a tight correlation of $L_{UV}$ with stellar mass does not necessarily mean a smooth star formation history. Rather, it just means that a significant amount of the variance in star formation is on time scales shorter than it takes for the UV-emitting stars to react. As we study fainter galaxies, we must seek to use star formation indicators that trace shorter times scales such as \Ha and other nebular emission lines.

We also emphasize that observationally a flux limited survey will be strongly biased towards galaxies in the burst phase. This is especially true when the selection is made on emission lines because the luminosities change dramatically. For example, Figure~\ref{fig:ms} shows that the \Ha selection of recent HST~WFC3/IR grism surveys (WISP, \citealt{atek10}; 3DHST, \citealt{brammer12}) can detect galaxies at $z\sim 1.4$ in a recent burst at stellar masses larger than $10^8 $~M$_{\odot}$, but will miss the vast majority of the galaxies at around $10^8 $~M$_{\odot}$. When the selection is made in the UV continuum the bias will be less severe since the UV scatter is lower but still significant. Figure 7 shows that the HUDF depths (\citealt{beckwith06}) can detect galaxies currently in a burst at $10^7$~M$_{\odot}$ but the depths need to be an order of magnitude lower to detect all galaxies of that mass. Indeed, UV selection of dwarf galaxies at these redshifts (\citealt{alavi14}) is still finding mostly galaxies which appear to be in a strong and recent burst of star formation (\citealt{stark14}).

Finally, we note that in Figure~\ref{fig:ms}, we calculate the average of the SFR and each observable for each galaxy by adding all of the individual values at each time step and dividing by the number of time steps (that is, to get a mean, rather than a geometric mean). The values are plotted with the large star symbols, and are analogous to what observers would obtain when stacking large samples of galaxies in bins of stellar mass. The slope of the \emph{average} SFR versus M$_{*}$ relation is near one, consistent with the slope measured by \citet{whitaker14} in galaxies with $9.3<\log_{10}({\rm M}_*/{\rm M}_{\odot})<10.2$ at these high redshifts. However, other authors such as \citet{henry13} find a shallower slope of $0.31\pm 0.08$ in galaxies with $8.2<\log_{10}({\rm M}_*/{\rm M}_{\odot})<9.8$. We believe this is a natural consequence of a bias toward galaxies in a burst phase when selecting via emission lines.  As a result, the lowest mass galaxies detected in that survey are those with the largest SFRs. Such a bias will naturally lead to perceiving a shallower main sequence.

\section{Comparison with physical properties from SED fitting}
\label{sec:sed}
In this section, we explore how well SED fitting to broadband photometry can determine the physical properties of bursty galaxies. In particular, we examine the effects of the incorrect parameterization of the SFH on determining the galaxies’ SFRs and stellar masses.

The code Fitting and Assessment of Synthetic Templates (\texttt{FAST}, see the appendix in \citealt{kriek09} for details) is used to find the best fitting SED template from a grid of BC03 models. This fitting procedure is applied to the model galaxies simulating the following photometry: the UV/optical channel (UVIS) of the Wide-field Camera 3 (WFC3) on board the Hubble Space Telescope (HST), the Advance Camera Survey (ACS)/WFC (F475W, F625W, F775W and F850LP) and the near-infrared (IR) with WFC3/IR (F125W and F160W). This is the same photometry used by \citet{alavi14} and it spans the rest-UV to rest-optical across the Balmer/4000\AA\ break. Nebular emission lines are not included in the photometry, nor are they considered in the SED templates. Therefore, we are isolating the effects of the bursty SFHs independent of complications due to emission line contributions to broadband photometry. A signal-to-noise of 20 is assumed for each photometric band, thus noise is not a significant issue. The IMF is assumed to be given by \citet{chabrier03}. The parameter space explored by \texttt{FAST} is given by $\log_{10}({\rm Age/{\rm Gyr}})=7$ up to a maximum age given by the age of the Universe, $A_{V}=0$ (since our model galaxies do not have any dust extinction), the metallicity is fixed to $Z=0.2 Z_{\odot}$ and the SFH $\propto \exp(+t/\tau)$ with $\tau=0.3$--10~Gyr. Note that we assume that, on average, the SFRs are exponentially rising. This has been suggested to be more accurate than exponentially decreasing SFHs by other authors such as \citet{reddy12} in describing star-forming galaxies at high redshifts. Then, physical properties are derived from these fits such as galaxy stellar mass and SFR.

\begin{figure}
\includegraphics[width=\columnwidth]{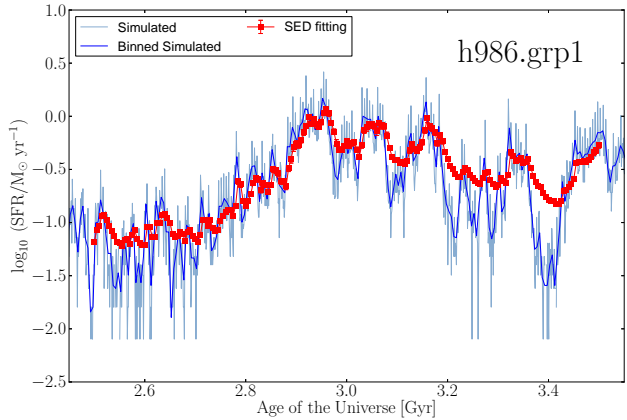}
\caption{The galaxy SFR versus time (when the Universe is between 2.5~Gyr and 3.5~Gyr old). The light blue line shows the simulated SFR with a resolution of 1~Myr, the dark blue line plots the simulated SFR averaged every 5~Myr. The red line shows the SFR estimated from the SED fitting every 5~Myr. At $z=2$, this galaxy has a stellar mass of $\log_{10}({\rm M}_{*}/{\rm M}_{\odot})=8.51$. Though the SED fitting reflects the average SFR, it can not recover the short timescale variations.}
\label{fig:h986}
\end{figure}

Figure~\ref{fig:h986} plots the galaxy SFR versus time for one of our simulated galaxies (h986.grp1, which has a stellar mass at $z=2$ of $\log_{10}({\rm M}_{*}/{\rm M}_{\odot})=8.51$). The figure is showing the simulated SFR with a 1~Myr resolution, the SFR from the simulations but averaged over 5~Myr, and the SFR that is estimated from the SED fitting. It is clearly illustrated that the SFR provided by the SED fitting does not react rapidly enough to the bursty variations of the actual SFR as a consequence of being based on the UV continuum. Furthermore, it is clear that the SFH parametrization does not reflect the variations on short time scales of the actual SFH. Yet, the average SFR on larger timescales is recovered.

\begin{figure}
\includegraphics[width=\columnwidth]{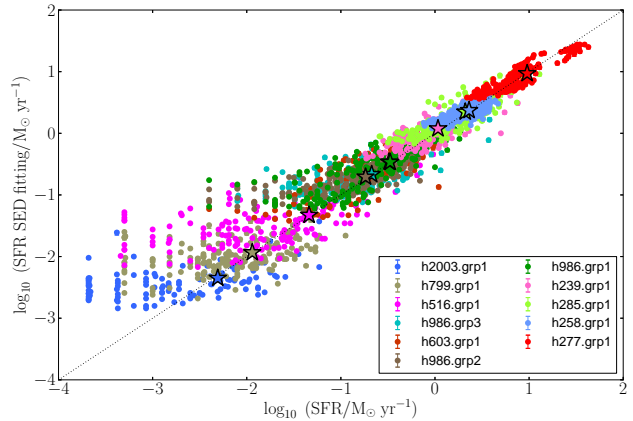}
\caption{The $x$-axis shows the model SFRs whereas the $y$-axis shows the SFRs estimated by the SED fitting. The stars are the mean for each simulated galaxy. Though the SED-derived SFR does not reflect the large dispersion in the galaxy SFR, the average values are recovered. The dotted line plots the 1:1 straight line to guide the eye.}
\label{fig:SFR}
\end{figure}

A comparison of the SFRs from the SED fitting and the simulated SFRs is shown in Figure~\ref{fig:SFR}. The average values are calculated by the linear (not geometric) mean.  This ensemble average is similar to what a deep-field survey would measure, as it randomly samples many galaxies at different stages of bursting and quenching to derive a star formation rate density. Interestingly, the average SED-derived SFRs agree well with the actual average SEDs on long time scales. That is, there is no bias in one direction. These conclusions apply only if we observe the entire population of galaxies at a given mass. If a flux-limited survey only selects the brighter galaxies at a given mass, then it will only select the galaxies in a new burst and the derived average SFR would be biased high. As expected, the actual SFR scatter matches the scatter derived by the SED fitting for the larger mass galaxies. However, as expected from the previous result shown in Figure~\ref{fig:h986}, the scatter in the SED-derived SFR is much lower than the actual SFR in the lower mass galaxies.

\begin{figure}
\includegraphics[width=\columnwidth]{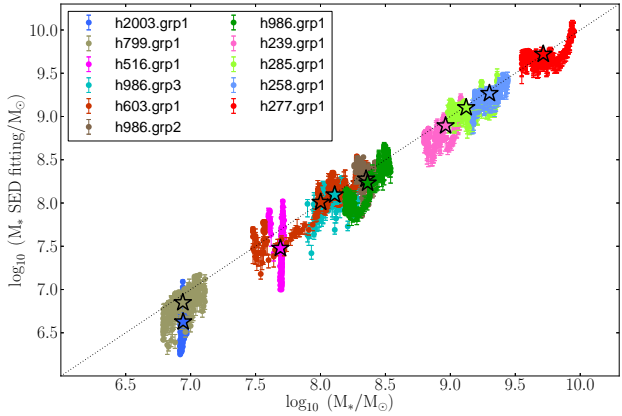}
\caption{The $x$-axis shows the model stellar masses whereas the $y$-axis shows the stellar masses estimated by the SED fitting. The stars are the geometric mean for each simulated galaxy. The SED-derived stellar masses recover the true stellar masses very well at $\log_{10}({\rm M}_{*}/{\rm M}_{\odot})> 8$. At lower masses, these are biased low because old stars can easily be hidden by the low mass-to-light ratio new burst of star formation. The dotted line plots the 1:1 straight line to guide the eye.}
\label{fig:mass}
\end{figure}

In Figure~\ref{fig:mass}, a comparison between the stellar masses from SED fitting and the stellar masses from the simulations is shown. The average values are calculated by the geometric mean of the stellar masses. On average, there is a good agreement for the larger mass galaxies but there is some bias for galaxies with $\log_{10}({\rm M}_{*}/{\rm M}_{\odot})\lesssim 8$. For these lower mass galaxies the SED-fitting underestimates the stellar mass by, on average, a factor of 0.2~dex. The reason is that recent bursts (with their high mass-to-light ratios) can completely dominate the high mass-to-light ratio light from older generations of stars. Qualitatively our results on the stellar masses are compatible with other analyses such as the one presented by \citet{mitchell13}. They conclude that stellar masses estimated from SED fitting may be under-estimated when considering dwarf galaxies that have undergone a recent burst of star formation. That is also true in our analysis.

\section{Summary}
\label{sec:summary}
The star formation histories of dwarf galaxies are thought to be characterized by short bursts followed by rapid quenching due to supernovae feedback. Typically, analysis of the broadband SEDs and nebular emission lines of high redshift galaxies does not consider star formation histories with short ($<10$~Myr) timescale variations. To the contrary, fits of the SFHs and ages are usually parametrized as slowly varying exponential functions (both decreasing and increasing). In this paper, we use the SFHs from numerical simulations of low mass galaxies with realistic prescriptions for supernovae feedback to consider several important consequences of \emph{bursty} star formation in these galaxies. Our galaxies are located at the redshift range where the cosmic star-formation rate peaks, namely $z\sim 2$--3, although similar principles apply to other redshifts. In particular, we analyse the $L_{\nu}(1500{\rm \AA})/L_{\nu}(900{\rm \AA})$ ratio and the \Lya and \Ha EWs. The star-forming main sequence is investigated as well in the stellar mass range of around 10$^{7}$--10$^{10}$~M$_{\odot}$. We stress that the values reported in this analysis are rest-frame and that dust extinction is not included. This fact will not change any of our conclusions. The main points presented in our analysis are briefly summarized in this section.

\begin{itemize}

\item Because the Lyman continuum, $L_{\nu}(900{\rm \AA})$, is emitted from only the most massive short-lived stars, it accurately traces the rapid and large variations in SFR in the low mass galaxies. Specifically, the LyC (and nebular emission lines related to the LyC) vary by 1.03, 0.45, and 0.22~dex in the mass ranges of $\log_{10}({\rm M}_{*}/{\rm M}_{\odot})<8$, $8 \leq \log_{10}({\rm M}_{*}/{\rm M}_{\odot})\leq 9$, and $\log_{10}({\rm M}_{*}/{\rm M}_{\odot})>9$, respectively.  Though the scatter in the non-ionizing UV continuum, $L_{\nu}(1500{\rm \AA})$, increases toward lower stellar mass, it does so at a much lower degree (0.43, 0.28, 0.16~dex in the same mass ranges). Finally, the continuum at optical wavelengths, $L_{\nu}(5500{\rm \AA})$, and longward increases by only approximately 0.05~dex from the most massive to least massive bins.  

\item Because the ionizing photons vary on different time scales than the longer wavelength continuum photons, the ratio of non-ionizing to ionizing continuum can vary significantly in low mass galaxies, as the SFR variation also increases. This ratio, $L_{\nu}(1500{\rm \AA})/L_{\nu}(900{\rm \AA})$ is an important number for converting direct LyC detections to LyC escape fractions or for converting observed UV luminosity functions to ionizing photon production rates. A constant value is often assumed. However, we show that this ratio can be lower by a factor of 2 in galaxies with recent bursts and can be very large after recent star-formation quenching. Careful consideration of this variation will need to be taken into account in future studies of low mass galaxies. 

\item Because of the widely varying luminosities of dwarf galaxies, it is very difficult to obtain complete samples at a given mass in flux-limited surveys. Selection of dwarf galaxies near the flux limit of a survey will be biased toward galaxies in the burst phase. This is of particular concern for H$\alpha$-selected samples of lower mass galaxies like those selected with WFC3/IR grism spectroscopy. Yet, it is also a concern for UV-selected dwarf galaxies as well. Such selection biases should be carefully considered when attempting to determine average properties of dwarf galaxies at high redshift in future studies. 

\item The ratio between UV and \Ha luminosity is a useful observable that quantifies burstiness in the SFH of dwarf galaxies. As new instruments allow \Ha detection in galaxies of lower mass at $z\sim 2$, this will be a particularly useful metric with which to quantify bursty star formation. Specifically, $\log_{10}[\nu L_{\nu}(1500{\rm \AA})/L_{{\rm H}\alpha}]>2.5$ signifies a recent shutdown of star formation on the timescale of less than 10~Myr. Determining this ratio in a large sample of dwarf galaxies will allow us to better constrain our models and simulations of supernovae feedback and bursty star formation.

\item Dwarf galaxies are only significantly producing ionizing photons in their burst phase. And because galaxies in this phase are preferentially selected (via UV continuum or nebular emission lines), it is much easier to find the galaxies responsible for the bulk of the ionizing emission than it is to detect a mass complete sample. Such selection biases and their effects on derived escape fractions or volume ionizing emissivities of dwarf galaxies should be carefully considered.  

\item While the scatter in $L_{\nu}(1500{\rm \AA})$ does increase towards lower stellar masses, the scatter in $L_{{\rm H}\alpha}$ is three times larger and closely reflects the actual variation in the SFR. Therefore, when characterizing the dispersion in SFR of galaxies with $\log_{10}({\rm M}_{*}/{\rm M}_{\odot})<9$, it is important to use a short timescale tracer of star formation such as H$\alpha$, and not the UV-to-optical continuum SEDs.

\item We investigate whether an incorrect (slowly varying) parametrization of the SFH results in incorrect derived quantities (\eg SFRs and stellar masses). We find that, on average, the SFRs derived from SED fitting are correct at all stellar masses. However, the SFR dispersion derived from SEDs is considerably smaller than the true SFR dispersion in these galaxies. These conclusions apply only if we sample the entire population of galaxies at a given mass, otherwise the SFR will be overestimated as flux-limited surveys will be biased toward selecting galaxies in a burst phase.

\item On average, for galaxies with $\log_{10}({\rm M}_{*}/{\rm M}_{\odot})> 8$, the stellar masses from SED fitting are correct. However, for galaxies with $\log_{10}({\rm M}_{*}/{\rm M}_{\odot}) < 8$, we find that stellar masses are underpredicted by the SED fitting by a factor 0.2~dex when assuming slowly varying SFHs.

\end{itemize}

%%%%%%%%%%%%%%%%%%%%%%%%%%%%%%%%%%%%%%%%%%%%%%%%%%%%%%%%%%%%%%%%%%%%%%%%%%%%%%
%%%%%%%%%%%%%%%%%%%%%%%%%%%%%%%%%%%%%%%%%%%%%%%%%%%%%%%%%%%%%%%%%%%%%%%%%%%%%%

\section*{Acknowledgements}
We are grateful to St\'ephane de Barros, Alaina Henry, Philip Hopkins, Jason X. Prochaska and Marc Rafelski for useful discussions. We also thank the referee for improving the paper. AMB acknowledges support from HST AR-12631, provided by NASA through a grant from the Space Telescope Science Institute, which is operated by the Association of Universities for Research in Astronomy, Incorporated, under NASA contract NAS5-26555.

\bibliographystyle{mn2e}

\label{lastpage}
\end{document}